\documentclass{aastex631}

\shorttitle{An H$\delta$ Absorption Line Selected Sample}
\shortauthors{Wu et al.}
\graphicspath{{./}{figures/}}

\begin{document}

\title{Starburst Galaxies in Their Last Billion Years: An H$\delta$ Absorption Line Selected Sample}

\author[0009-0008-9984-1681]{Shumei Wu}
\affiliation{Chinese Academy of Sciences South America Center for Astronomy, National Astronomical Observatories, CAS, Beijing 100101, People’s Republic of China}
\affiliation{CAS Key Laboratory of Optical Astronomy, National Astronomical Observatories, Chinese Academy of Sciences, Beijing 100101, People’s Republic of China}

\author[0000-0001-6511-8745]{Jia-Sheng Huang}
\affiliation{Chinese Academy of Sciences South America Center for Astronomy, National Astronomical Observatories, CAS, Beijing 100101, People’s Republic of China}
\affiliation{CAS Key Laboratory of Optical Astronomy, National Astronomical Observatories, Chinese Academy of Sciences, Beijing 100101, People’s Republic of China}
\affiliation{Harvard-Smithsonian Center for Astrophysics, 60 Garden Street, Cambridge, MA 02138, USA}

\author[0000-0003-0202-0534]{Cheng Cheng}
\affiliation{Chinese Academy of Sciences South America Center for Astronomy, National Astronomical Observatories, CAS, Beijing 100101, People’s Republic of China}
\affiliation{CAS Key Laboratory of Optical Astronomy, National Astronomical Observatories, Chinese Academy of Sciences, Beijing 100101, People’s Republic of China}

\author[0000-0002-7928-416X]{Y. Sophia Dai}
\affiliation{Chinese Academy of Sciences South America Center for Astronomy, National Astronomical Observatories, CAS, Beijing 100101, People’s Republic of China}
\affiliation{CAS Key Laboratory of Optical Astronomy, National Astronomical Observatories, Chinese Academy of Sciences, Beijing 100101, People’s Republic of China}



\begin{abstract}
In this paper, we focus on the study of starburst galaxies in their final billion years. Our galaxy selection is based solely on the presence of the H$\delta$ absorption line, which permits tracing the later evolution of starburst galaxies, coinciding with the emergence of A-type stars in these galaxies. We propose a novel method that utilizes star formation rate and UVJ colors to classify galaxies in the sample, and use the spectral features to mark their evolution stages.  Our in-depth analysis of the MgII line indicates the substantial increasing of F- and G-type stars when a galaxy evolves from star forming to quiescent phase. Furthermore, we identify AGNs in this sample to explore their roles in the later stage of galaxy star formation history.
\end{abstract}



\section{Introduction} \label{sec:intro}


Galaxy mass assembly occurred through their star formation activities over the course of cosmic time. However, observations \citep{2007ApJ...665..265F, 2013A&A...556A..55I, 2013ApJ...766...21H,2020ApJ...905..103X} have revealed a drastic decrease in the number of massive galaxies in their mass functions, suggesting that star formation cannot be sustained indefinitely. This evolutionary trend is also evident in the galaxy color bimodality observed in the color-mass diagram \citep{2004ApJ...600..681B,2024ApJ...961..216C}. Star-forming galaxies with low mass initially reside in the blue cloud region and then transition across the green valley to become passive galaxies in the red sequence region \citep{2004A&A...424...23F, 2004AJ....128.2652G, 2007ApJ...669..184A,2007A&A...476..137A}. During the transition through the green valley, star formation quenching is believed to occur. Several quenching scenarios have been proposed, including AGN feedback \citep{1998A&A...331L...1S, 1999MNRAS.308L..39F, 2001ApJ...551..131C, 2005ApJ...630..705H, 2006ApJS..163...50H, 2006MNRAS.365...11C, 2009ApJ...699...89C, 2012ApJ...754..125C}, halo heating and stripping \citep{2021ApJ...907..114D, 2023ApJ...949...94G}, and even galaxy gas starvation \citep{2013ApJ...775..124F}. However, observational studies of these quenching mechanisms are less conclusive.

 
Galaxies in the transition zone, known as the green valley, must undergo the quenching of star formation before they enter the red sequence area and become passive galaxies. Cosmological star formation density peaked at $z \sim 2$ suggests that there would be a large sample of post-starburst to quiescent galaxies at intermediate redshifts ($z<2$). Although post-starburst galaxies also bridge the gap between starburst and quiescent galaxies, the relationship between the green valley and post-starburst galaxies has yet to be established. On the other hand, when the star formation history of a galaxy is complex, its stellar populations are often mixed, making the selection of post-starburst to quiescent galaxies less effective. Since optical spectral features strongly depend on the stellar population, a large spectroscopic sample is still required to statistically understand how post-starburst galaxies evolve into the quiescent phase.

The Balmer absorption lines can provide insight into the evolutionary path of how a galaxy transitions across the green valley. The absorption lines could clearly exhibit the information about stellar age, star formation history, and more, which are highly informative for understanding the galaxy quenching process. The strong Balmer absorption line indicates the domination of A-type stars, representing the post-starburst phase. Thus one approach to studying galaxy population evolution is to examine elliptical galaxies with A-type star spectral features, known as E+A galaxies \citep{1983ApJ...270....7D}. Since the Balmer lines in young galaxies primarily appear as emission lines, a sample selected based on Balmer absorption lines would encompass galaxy populations ranging from post-starburst to quiescent. Wide-field surveys such as the SDSS have enabled several groups \citep{1996ApJ...466..104Z, 2004MNRAS.355..713B, 2007MNRAS.381..187G, 2008ApJ...688..945Y} to assemble large samples of E+A galaxies. However, the sensitivity of the SDSS spectrum limits the H$\delta$ absorption-selected sample to low redshift galaxies ($z<0.3$). 

In this project, we are studying a spectroscopic sample of galaxies selected based on the presence of the Balmer H$\delta$ absorption line. This selection allows us to study galaxies with existing A-type or later stage stellar population, which typically last for a couple of billion years depending on the galaxy's star formation history. Our focus is on the late-stage evolution of starburst galaxies, starting from the appearance of A-type stars to quiescent galaxies, rather than solely post-starburst galaxies. Our sample was selected from the AGN and Galaxy Evolution Survey (AGES) spectroscopic sample \citep{2012ApJS..200....8K}. The AGES sample for galaxy spectroscopy was completed based on their I and [3.6] magnitudes, with a range up to I=20. The survey also includes a faint AGN sample with magnitudes between 20 and 22. We specifically selected galaxies with the Balmer H$\delta$ absorption line from the AGES sample, aiming to obtain a galaxy sample representing the late stage of the star-forming era and the early quenching process. Our sample covers a redshift range of 0.1$<z<$0.9, which corresponds to the decreasing slope of the cosmological star formation density since $z=2$. At redshifts greater than 1, the H$\delta$ line shifts to the near-infrared bands. So our sample reaches the highest redshift range for optical spectral surveys and covers an area of approximately 8 square degrees. 
Additionally, \citet{2009ApJ...707.1387C} presented an optical spectroscopic sample for a Spitzer/MIPS 24$\mu$m selected sample, including a substantial number of galaxies with the Balmer absorption line in a redshift range similar to ours, although their sample area coverage is four times smaller.



 The structure of this paper is as follows. Section 2 details our sample selection process. In Section 3, we present the physical properties of the sample and analyze its evolutionary history. The scientific implications of this sample are discussed in Section 4, 5, 6, and we conclude in Section 7. Throughout the paper, we assume a flat CDM cosmology with $\Omega_M= 0.3$, $\Omega_\Lambda= 0.7$, and $h=0.7$, and we adopt the Chabrier initial mass function \citep[IMF,][]{2003PASP..115..763C} for stellar population modeling.
 
\section{The absorption line Sample Selection} \label{sec:sample}

We selected our sample from the AGES spectroscopic sample with the strong Balmer absorption lines. We follow previous studies to use only H$\delta$(4102\AA).  The advantage of H$\delta$ is that its emission line is weak, and there are no other emission lines nearby. This makes it robust to measure the equivalent width. 

\subsection{The Photometric and Spectral Samples in the Bo\"{o}tes Field} \label{subsec:spec}

The full-band photometric data in the Bo\"{o}te field spans from X-ray to radio bands. The optical band data (Bw, R, I) are sourced from the NOAO Deep Wide-Field Survey \citep[NDWFS, ][]{1999ASPC..191..111J}. Near-infrared band data (J, Ks) are obtained from FLAMINGOS Extragalactic Survey \citep[FLAMEX, ][]{2006ApJ...639..816E}. Mid-infrared band data (3.6, 4.5, 5.8, 8.0 $\mu$m) is sourced from the IRAC Shallow Survey \citep{2004ApJS..154...48E}, with 24$\mu$m data from Soifer \& Spitzer/NOAO Team \citep{2004AAS...204.4805S}. 
UV band data are sourced from GALEX \citep{2005ApJ...619L...1M}, and X-ray band data is from the Chandra XBo\"{o}tes survey \citep{2005ApJS..161....1M,2005ApJS..161....9K,2006ApJ...641..140B}.


The AGN and Galaxy Evolution Survey \citep[AGES,][]{2012ApJS..200....8K} is a spectral survey conducted in the Bo\"{o}tes field. Spectroscopic observations were carried out using Hectospec on the MMT 
\citep{2005PASP..117.1411F}. The survey consists of a total of 15 Hectospec pointings, each covering 1 deg$^2$, to encompass a total area of 7.9 deg$^2$. The spectral wavelength range is 3700\AA\ - 9200\AA\ with a 1.2\AA\ pixel scale and the spectral resolution is R $\sim$ 1000. The AGES targets were selected based on multi-wavelength photometric data, with the sampling rate determined using I and IRAC [3.6] magnitudes up to I$=$20. The sample comprises a total of 21,462 spectra, with a redshift range of 0.1$<z<$1 \citep{2005PASP..117.1411F}. The final redshift identification rate is reported to be between 85-97$\%$ in each Hectospec field of view.

The AGES spectral wavelength range spans from 3700 to 9200\AA, allowing for the detection of H$\delta$ up to $z<1.2$. However, the sensitivity of AGES spectroscopy for absorption lines diminishes at observed wavelengths longer than 7500\AA\ due to high sky background. For our specific study, we selected our sample in the redshift range of $0.1<z<0.8$ to ensure high signal-to-noise measurements of absorption lines. The volume at $z<0.1$ for our sample is too small to include any galaxies.

\subsection{H\texorpdfstring{$\delta$}{} Absorption Line Measurement and Sample Selection} \label{subsec:mir}

\begin{figure*}[ht!]
\centering
\includegraphics[width=0.85\linewidth,clip=true, trim=12 12 0 0]{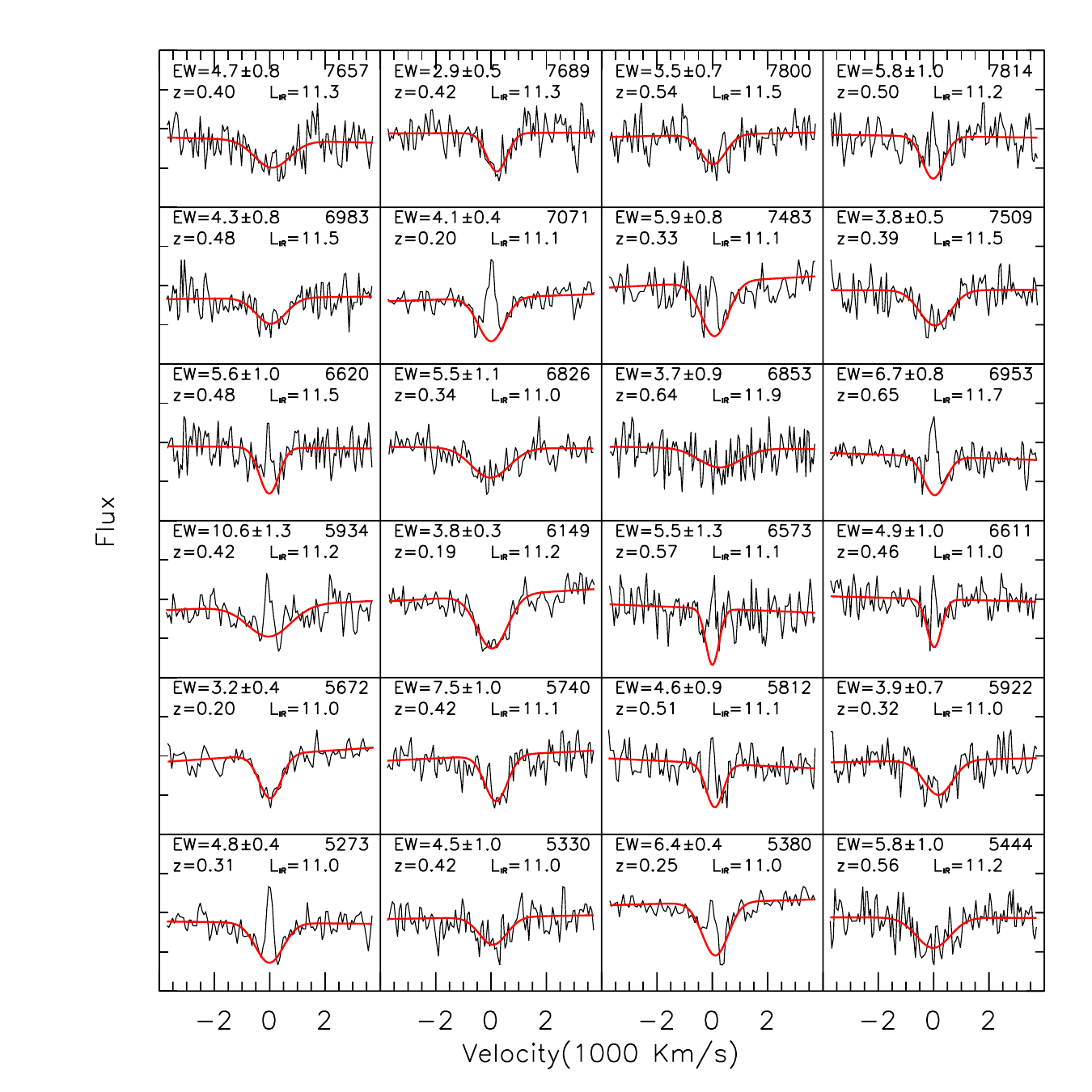}
\caption{Examples of our absorption line fitting. The red line model profile represents the final fitting result after 3 iterations. The resulting equivalent width and its redshift are displayed in the upper-left corner, while the object ID is shown in the upper-right corner in each panel.}
\label{fig:fig1}
\end{figure*}

We measured the H$\delta$ absorption line equivalent width for spectra within the AGES sample, specifically focusing on the redshift range of 0.1$<z<$0.8. The H$\delta$ absorption line equivalent width is defined as follows: the nearby continuum is measured in the wavelength range of 4041.6$<\lambda<$4079.75\AA\ and 4128.5$<\lambda<$4161.0\AA, while the absorption line profile is measured in the range of 4083.5$<\lambda<$4122.25\AA. It's important to note that some galaxies also exhibit H$\delta$ emission lines in the absorption lines. To address this, we initially fitted the entire profile with a two-Gaussian model, encompassing one emission and one absorption line. We then subtracted the emission line component to isolate and fit the absorption line, and subsequently subtracted the updated absorption to measure the emission line profile. This process was iterated three times to obtain a stable measurement for both components. In the majority of cases within the sample, the absorption line components are significantly wider than their emission counterparts, ensuring a robust measurement of the absorption line equivalent width. However, a few galaxies with much stronger emission were deemed to have unreliable absorption line measurements and were excluded from our analysis. Our fitting method also provides an error estimate for the absorption line equivalent width. Figure~\ref{fig:fig1} displays selected examples of our line profile fitting.

\begin{figure*}[ht!]
\centering
\includegraphics[width=0.6\linewidth,clip=true, trim=12 12 0 0]{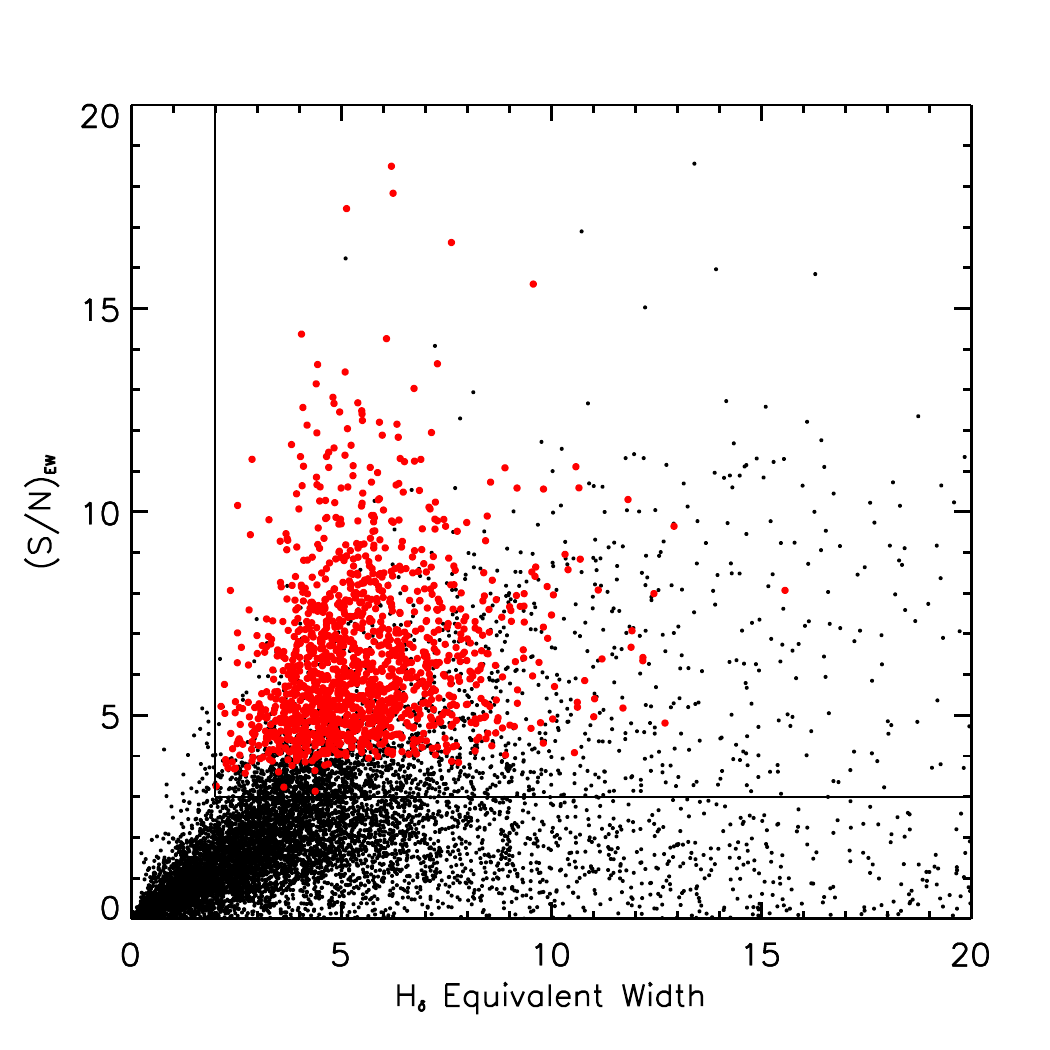}
\caption{Our sample selection. The black dots represent the measured equivalent width and their errors of H$\delta$ for the entire AGES spectral sample. The red dots indicate our selected sample. Both straight lines represent the two additional selection criteria as discussed in the text. There is a substantial number of galaxies with larger EW(H$\delta$) and (S/N)$_{EW}>$3, which are not selected into the sample, due to their lower continuum ($<5\sigma$) in both the ranges of 4041.6$<\lambda<$4079.75\AA\ and 4128.5$<\lambda<$4161.0\AA.}
\label{fig:fig2}
\end{figure*}

The sample selection criteria primarily rely on the signal-to-noise ratio for the continuum around H$\delta$.  Reliable measurement of the equivalent width of H$\delta$ absorption lines is dependent on both continuum and absorption line. We use the H$\delta$ equivalent width (EW) and its uncertainty $\sigma_{\rm EW}$ in each spectrum for the sample selection. This will enhance detectability of the absorption line. Our sample selection criteria are
\begin{equation}
    EW(H\delta)>2\text{\AA};
\end{equation}
\begin{equation}
    S/N=EW(H\delta)/\sigma_{\rm EW}>3.
\end{equation}
Both criteria provide reliable estimates of the EW for the sample and effectively reject false measurements caused by noise spikes. However, a low significance in the continuum can still lead to a large uncertainty in the EW estimation. We further require a 5$\sigma$ significance for the continuum in both the ranges of 4041.6$<\lambda<$4079.75\AA\ and 4128.5$<\lambda<$4161.0\AA\ to ensure a reliable measurement of the EW(H$\delta$). Figure~\ref{fig:fig2}  illustrates our sample in the $EW$-$(S/N)_{\rm EW}$ space, comparing the same measurement for the entire AGES spectroscopic sample. Our selection process rejects a substantial number of very large equivalent width measurements due to their very low continuum.With these criteria, we selected a total of 1323 galaxies with H$\delta$ absorption lines. Line profiles for a random subsample are demonstrated as examples in Figure~\ref{fig:fig1}. 

Our selection of high significance of continuum in wavelength range of 4041-4161\AA\ biases the sample towards blue-band luminous galaxies. Within this sample, 519 galaxies exhibit both H$\delta$ emission and absorption lines, while 804 galaxies only display H$\delta$ absorption lines. The full width at half maximum (FWHM) for all measured absorption lines in the sample is at least a factor of 2 wider than those for the measured emission lines. The mean absorption-emission line width ratio is approximately 5, indicating the robustness of our absorption line profile measurement for galaxies with both absorption and emission lines.

 

\section{Properties of the H\texorpdfstring{$\delta$}{} absorption line selected Sample}
\subsection{Stellar Mass and Star Formation Rate Estimation} 
\label{subsec:sfr}


The multi-wavelength photometry in the Bo\"{o}tes field allows for the estimation of stellar mass in our sample. We employed {\sc fastpp}\footnote{https://github.com/cschreib/fastpp} \citep{2009ApJ...700..221K} with stellar population model \citep{2003MNRAS.344.1000B} and  \citet{2000ApJ...533..682C} dust extinction
law ($0 < \rm Av < 3.$) to fit the measured spectral energy distributions (SEDs) of our sample. The selection effect of this sample is too complex to permit completeness correction and yield a mass function.  The I-band selection for AGES spectroscopy sets a low mass limit for this sample. 

\begin{figure*}[ht!]
\centering
\includegraphics[width=0.6\linewidth,clip=true, trim=12 12 0 0]{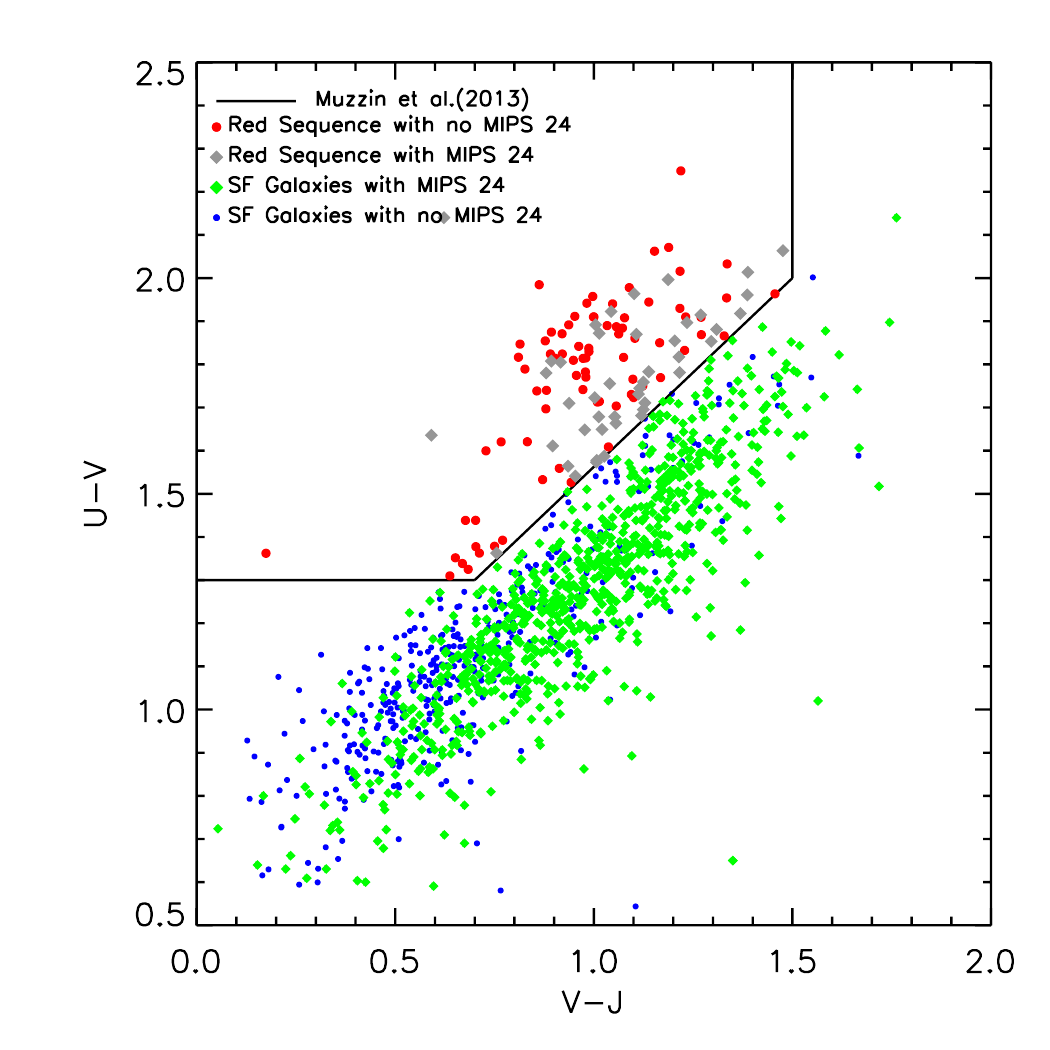}
\caption{Our sample in the UVJ diagram. The solid lines, taken from \citep{2013ApJ...777...18M}, divide the targets into passive galaxies (upper left region) and star-forming galaxies (remaining region).
123 galaxies, including 44 MIPS 24$\mu$m sources, are located in the passive galaxy area in this diagram.\label{fig:fig3}}
\end{figure*}

We divided our sample into four populations according to their UVJ colors \citep{2013ApJ...777...18M} and MIPS detection. Galaxies were classified as red sequence galaxies 
or star forming galaxies 
based on their location in the UVJ diagram (Figure~\ref{fig:fig3}). 
In each region galaxies were further divided into two population: those with MIPS 24 $\mu$m detection and without the detection.  Numbers of galaxies in each population are provided in Table~\ref{tab:1}.
There are 123 galaxies in the passive galaxy area, as shown in Figure~\ref{fig:fig3}, which accounts for roughly 10\% of the total sample.
\begin{deluxetable}{lll}
\linespread{1}
\tablewidth{500pt} 
\tablenum{1}
\tablecaption{Galaxy populations in the Sample. \label{tab:1}}
\startdata 
                &   &  \\
 red sequence without MIPS 24 detection & 79 &  \\ 
 red sequence with MIPS 24 detection & 44 &  \\  
 star forming galaxies with MIPS 24 detection & 801 & \\ 
 star forming galaxies without MIPS 24 detection & 399 & \\
\enddata
\end{deluxetable}

The MIPS observation of Bo\"{o}tes field was conducted in the Spitzer GTO survey \citep{2021A&A...648A...3K}. The limiting flux density at the 24$\mu$m channel is 150$\mu$Jy. There are 845 galaxies in the sample with MIPS 24$\mu$m detection, of which 801 galaxies were detected in at least one Herschel SPIRE band. The Herschel SPIRE flux densities were measured using the MIPS 24 $\mu$m image as a prior \citep{2021A&A...648A...3K}. 
We fitted the FIR SED templates of \citet{2001ApJ...556..562C} to the IR photometry to estimate their total IR luminosity. The fitting yields 10 ULIRGs, 361 LIRGs, and 415 galaxies with 10$<\log_{10}(L_{\rm IR}/L_{\odot})<$11.

\begin{figure*}[ht!]
\centering
\includegraphics[width=0.6\linewidth,clip=true, trim=12 12 0 0]{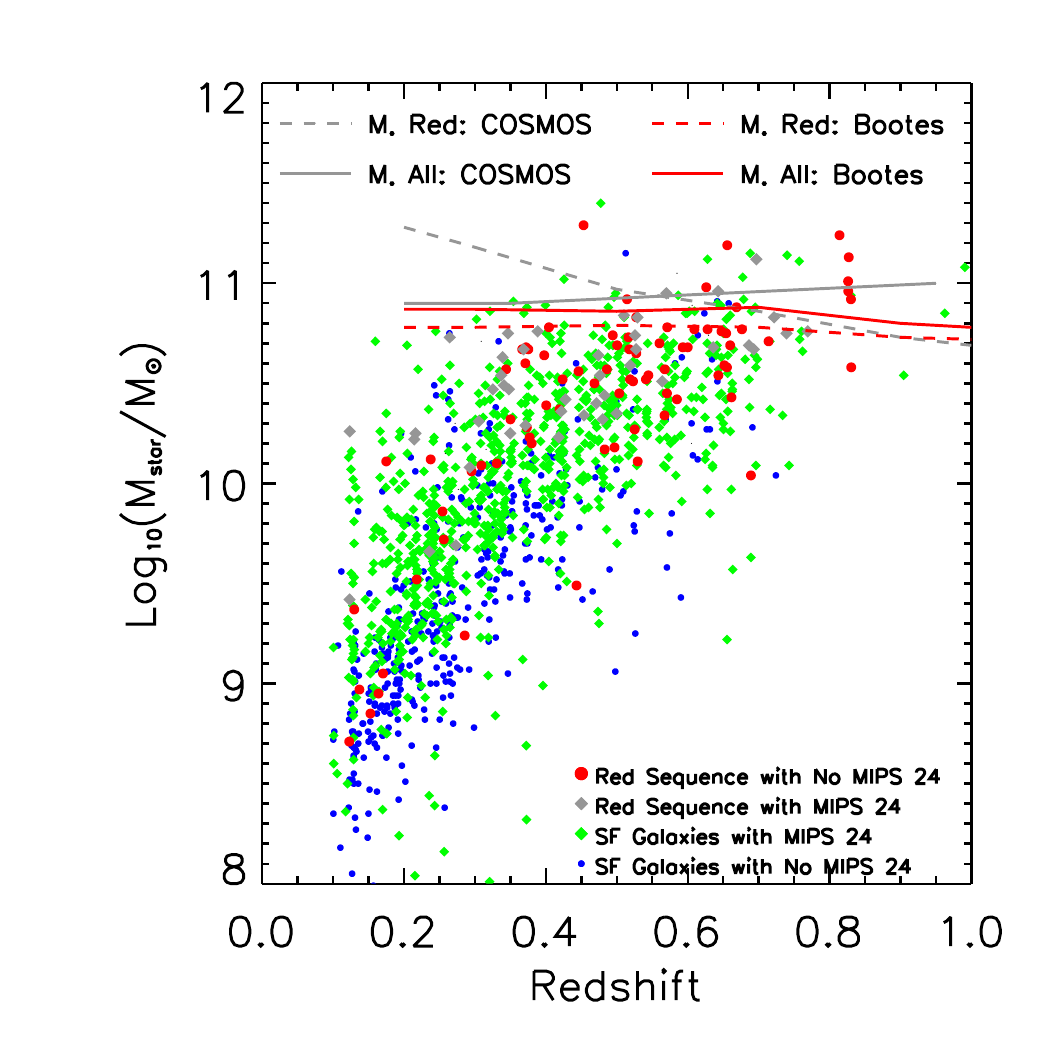}
\caption{Stellar masses of our sample are plotted against their redshifts. 
We also plot the characteristic masses (M$_{\bullet}$), which are parameters of Schechter functions fitting to galaxy samples, for red galaxies and the entire galaxy sample in the COSMOS field \citep{2010ApJ...709..644I, 2023A&A...677A.184W} and Bo\"{o}tes field \citep{2015ApJ...815...94B,2019ApJ...873...78B} for comparison. 
\label{fig:fig4}}
\end{figure*}


Most of the stellar masses in this sample are lower than the characterized masses of the mass functions in the Bo\"{o}tes field \citep{2015ApJ...815...94B,2019ApJ...873...78B}, as shown in Figure~\ref{fig:fig4}. The majority of galaxies identified as passive galaxies fall within the massive range of 10 $<\log_{10}(M_{\rm star}/M_{\odot})<$ 11. There are only a few galaxies with masses larger than the characterized masses of $\log_{10}(M_{\rm star}/M_{\odot})\sim$10.8. In the down-sizing scenario \citep{1996AJ....112..839C}, massive galaxies have already become passive galaxies at $ z<1$ and would not be selected in our sample.

\begin{figure}
\plotone{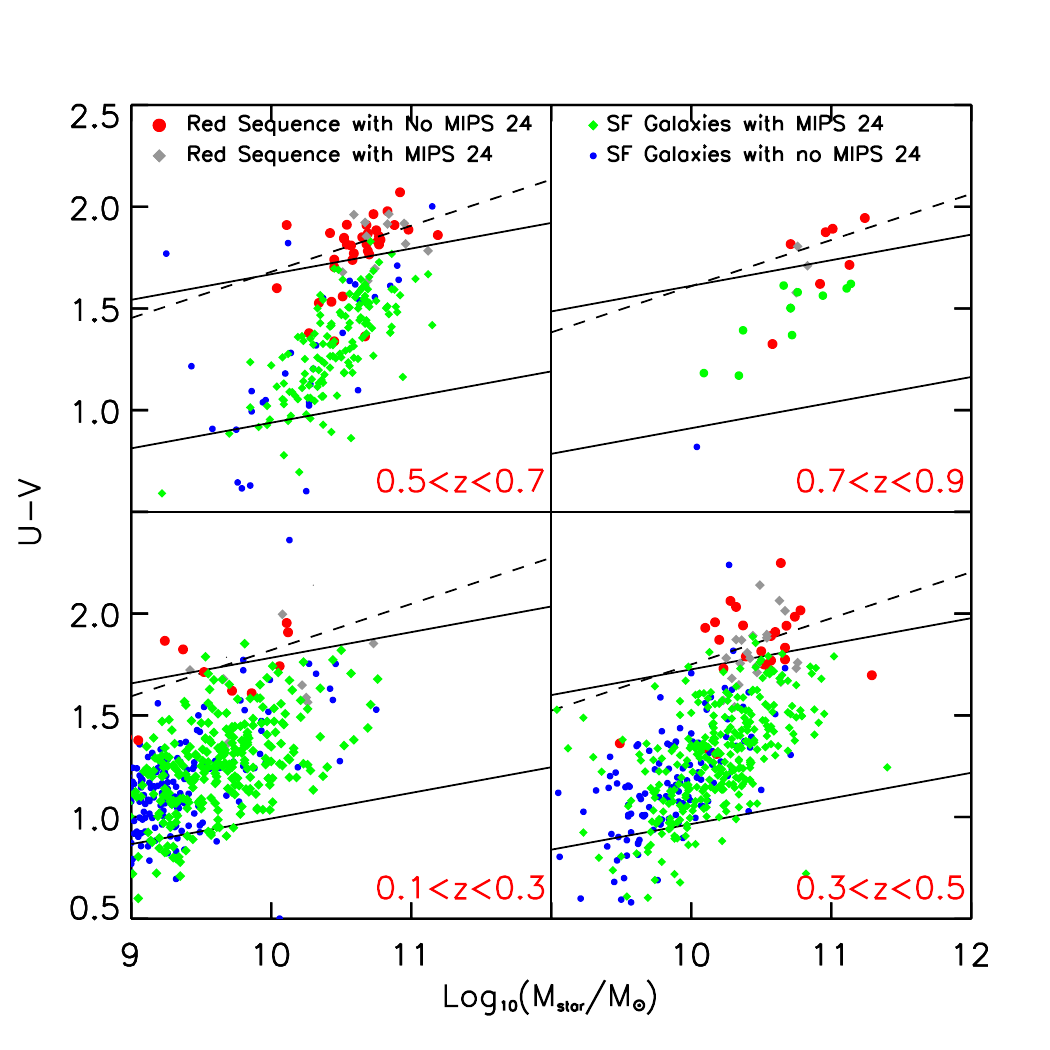}
\caption{The color–mass distribution for galaxies in our sample in 4 redshift bins of $z=$ 0.2, 0.4, 0.6 and 0.8. The dashed line represents the criterion separating galaxies in the red sequence and blue cloud, adopted from \citet{2006A&A...453..869B}. Both solid lines define the green valley, as given by \citet{2017A&A...601A..63W}.\label{fig:fig5}}
\end{figure}

Our UVJ classification for this sample is broadly consistent with their locations in the mass-color diagram. In Figure~\ref{fig:fig5}, most passive galaxies identified in the UVJ diagram appear in the red sequence region. Most star-forming galaxies in our sample are in the green valley region. Galaxies in the upper part of the green valley have the same mass range as those in the red sequence, implying a direct evolutionary link between these two populations. In a flux-limited sample, some galaxies in the red sequence are more massive than any of the galaxies in the green valley \citep{2006ApJ...641L..93F}. This suggests that galaxy merging is required to assemble more mass to become a massive red galaxy.

It is evident that there were star formation activities in most galaxies of this sample, as these galaxies were detected either in the rest-frame UV or MIPS 24 $\mu$m band. We estimated the total star formation rate for galaxies in the sample using both UV and IR photometry to combine both unobscured and obscured parts, as described in \citet{2014ApJ...795..104W}. The conversion is based on \citet{2005ApJ...625...23B} with the Chabrier IMF, as follows:

\begin{equation}
 {\rm SFR}[M_\sun  {\rm yr}^{-1}] = 1.09 \times 10^{-10} \times 2.2 L_{\rm UV} [L_\sun]+ {\rm SFR}_{\rm IR}[M_\sun  {\rm yr}^{-1}],
\end{equation}
where $L_{\rm UV}$ is the total integrated rest-frame luminosity at 1216–3000\AA. We estimated $L_{\rm UV}$ by the rest-frame continuum luminosity at 2800\AA, $L_{\rm UV} = 1.5 \times \nu L_{\nu,2800}$, and used the Spitzer/MIPS 24$\mu$m data to calculate the obscured SFR$_{\rm IR}$ \citep{2009ApJ...692..556R}. For the galaxies without the 24$\mu$m detection, we corrected for dust attenuation \citep{2000ApJ...533..682C} to yield a total star formation rate of SFR$_{\rm UV,corr}$\citep{2022ApJ...936...47C,2020ARA&A..58..529S}:
\begin{equation}
  {\rm SFR}_{\rm UV,corr}[M_\sun  {\rm yr}^{-1}] = {\rm SFR_{UV}} \times 10^{0.4 \times 1.8 \times{\rm  A_V}} .
\end{equation}

\begin{figure}[ht]
\centering
\includegraphics[width=0.495\linewidth]{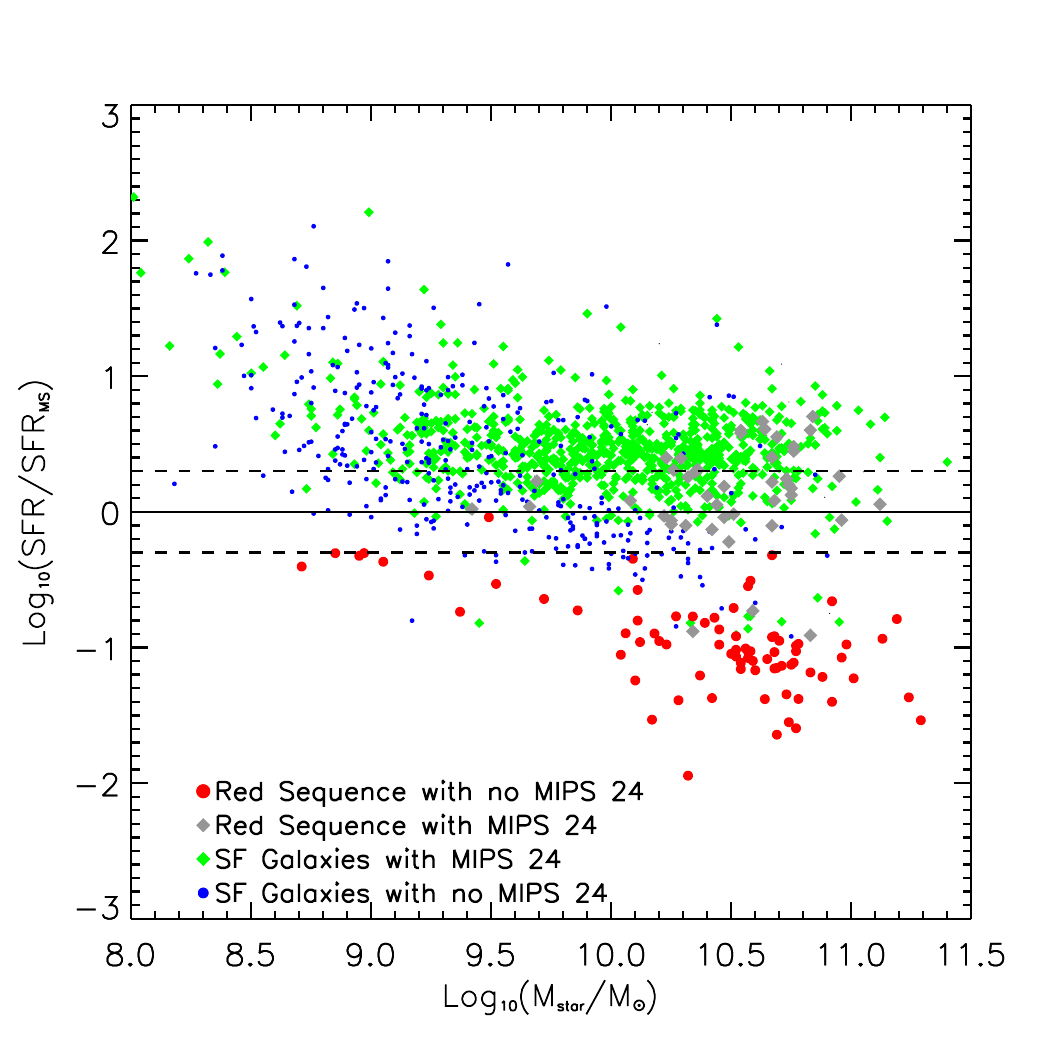}
\includegraphics[width=0.495\linewidth]{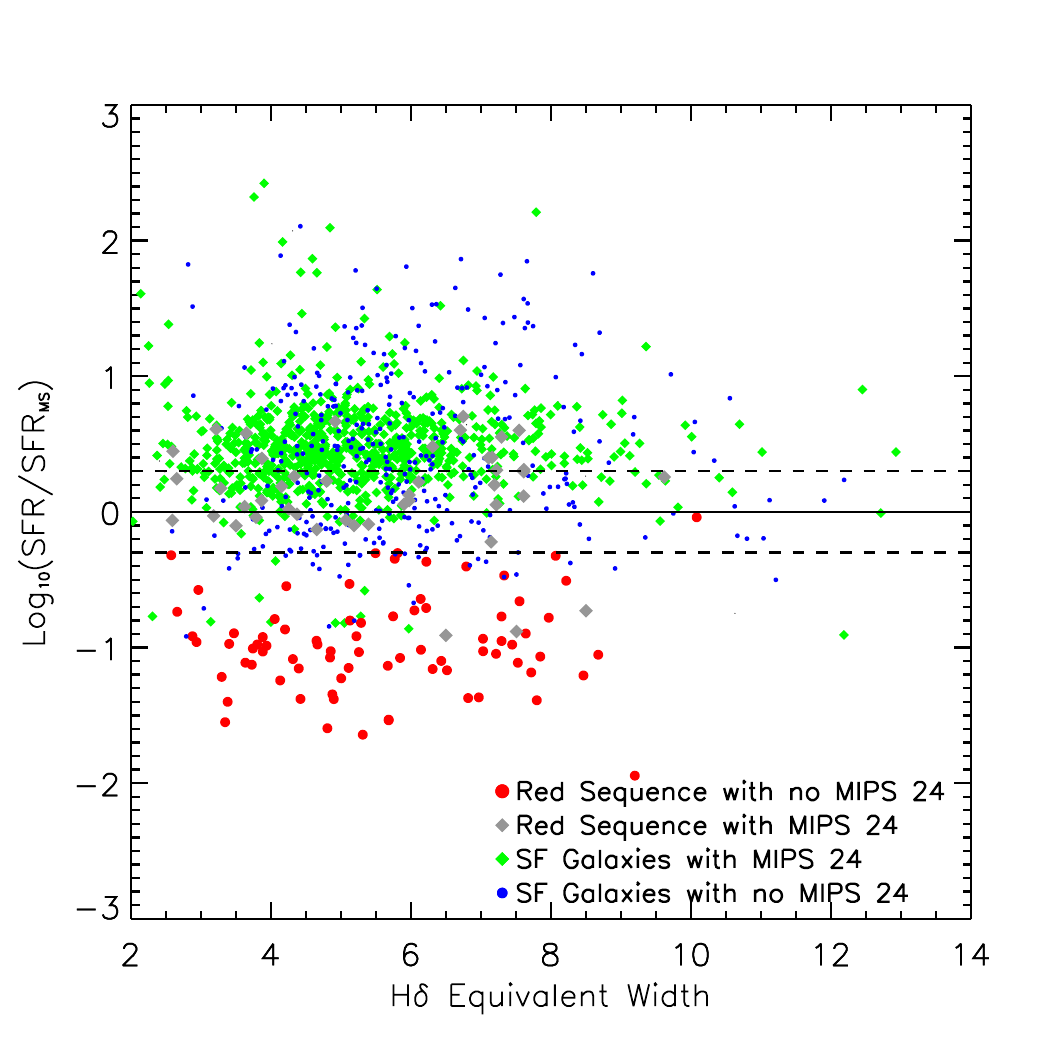}
\caption{Stellar mass ({\bf left panel}) or H$\delta$ Equivalent width ({\bf right panel}) versus star formation rate normalized by the model star formation rate for main sequence galaxies at the same redshift \citep{2019MNRAS.483.3213P}.  We also plot the main sequence as the solid line, and dashed lines 0.3 dex above and below the main sequence line to mark the main sequence area. Galaxies above the 0.3 dex line are starburst galaxies, and those below the -0.3 line are quiescent galaxies. Both starburst and quiescent galaxies have lower H$\delta$ equivalent width in the right panel.
}
\label{fig:fig6}
\end{figure}

The range of star formation rates is 0.1$<$SFR$<100 \, M_{\odot}\, \rm yr^{-1}$, with a mean of 1 $M_{\odot} \, \rm yr^{-1}$. To enable a fair comparison, we normalized the star formation rate for each galaxy to its model star formation rate for its main sequence counterpart \citep{2019MNRAS.483.3213P}. We studied the star formation status in our sample in the mass vs SFR/SFR$_{\rm MS}$ diagram (Figure~\ref{fig:fig6} left panel), and found that 852(64\%) galaxies in the sample are starburst galaxies with log$_{10}$(SFR/SFR$_{\rm MS}$)$>$0.3, 350(26\%) galaxies are in the main sequence with -0.3$<$log$_{10}$(SFR/SFR$_{\rm MS}$)$<$0.3, and 121(9\%) galaxies are below the main sequence. This is contrary to a normal flux-limited sample in which most galaxies are in the main sequence. Galaxies in the main sequence may be just on an evolutionary track moving from the starburst region to the passive galaxy region. There are star forming galaxies under the main sequence as shown in Figure~\ref{fig:fig6}. We will further investigate galaxies passing the main sequence line using H$\delta$.

\section{Tracing Galaxy Evolution with H\texorpdfstring{$\delta$}{} Absorption Line}\label{sec:Evolution}


Galaxies selected with H$\delta$ have an evolutionary history over a rather short time span. An A-type star has a lifetime of 1 billion years. \citet{2003MNRAS.341...33K} built a simple H$\delta$ absorption equivalent width model for a star-forming galaxy. Their model predicted that a galaxy started with EW(H$\delta$)=4\AA\ at the early star formation peak, reached a maximum of EW(H$\delta$) at an age of 4$\times10^8$yr, and then decreased to EW(H$\delta$)=2\AA\ at an age of $\sim10^9$yr. Therefore, our sample covers an evolutionary time span of roughly $10^9$ years.


We studied how the star formation rate changes with the A-type star population in the last 1 billion years of its starburst stage. In Figure~\ref{fig:fig6} right panel, both starburst galaxies with intensive star formation and quiescent galaxies have rather small H$\delta$ equivalent width. Those with larger H$\delta$ equivalent width EW(H$\delta$)$>8$\AA\ have a relatively low star formation rate and lie in or close to the main sequence area. This is consistent with the model of \citet{2003MNRAS.341...33K}, where the H$\delta$ equivalent width is lower in the early and late stages of a galaxy's star formation history.

\begin{figure}[ht]
\centering
\includegraphics[width=0.8\linewidth]{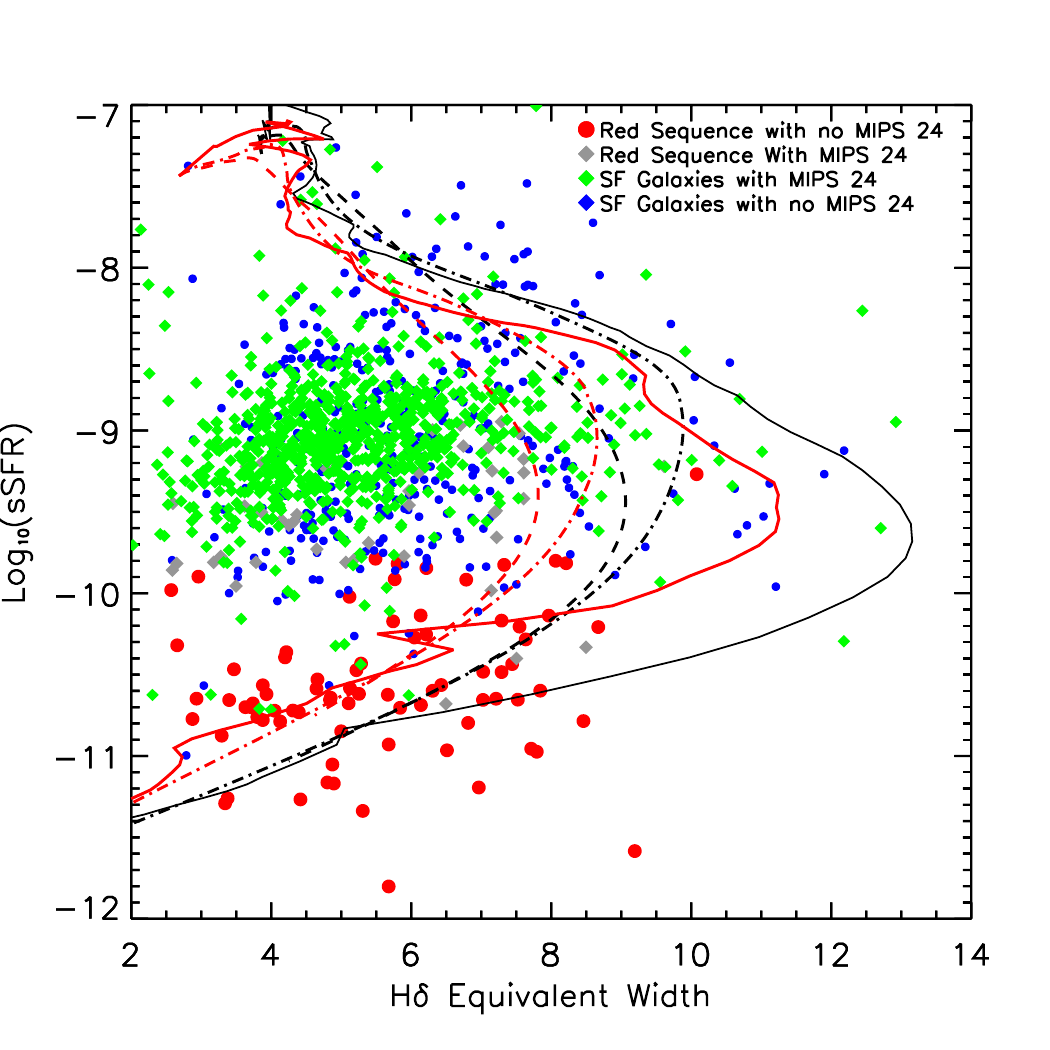}
\caption{H$\delta$ Equivalent width versus specific star formation rate. Both grey and red dots are those identified as passive galaxies in the UVJ diagram of Figure~\ref{fig:fig3}. The solid curve line represents the simple stellar population model, while the dash-dotted and dashed lines represent exponential decay models with time scales of 10$^6$ and 10$^7$ years. The black curves are models with solar metallicity, and the red curves are models with subsolar metallicity. The star formation rate in the models is also derived with $L_{2800}$ measured in the stellar population SED models, consistent with our sample measurement.}
\label{fig:fig7}
\end{figure}

Tracks of galaxy evolution across the main sequence line depend on galaxy star formation history. We constructed a set of simple models for specific star formation rates as a function of H$\delta$ equivalent width in Figure~\ref{fig:fig7}. The star formation histories in our models include single stellar population, exponential decay with $\tau$=10$^6$ and 10$^7$ years. We aim to use these models tracing evolution path for those passive galaxies. Most star forming galaxies in the sample may have very complicated  star formation history. For example, galaxy merging can move them up to the starburst region again in Figure~\ref{fig:fig7}.Passive galaxies may already ceased their star formation, their evolution track can be very similar to these models. Our models show that H$\delta$ equivalent width increases with decreasing star formation rate, reaches its peak in the main sequence region, and then decreases to the red sequence.

\section{Post Starburst Galaxies} \label{sec:psb}

\begin{figure}
\centering
\includegraphics[width=0.7\linewidth]{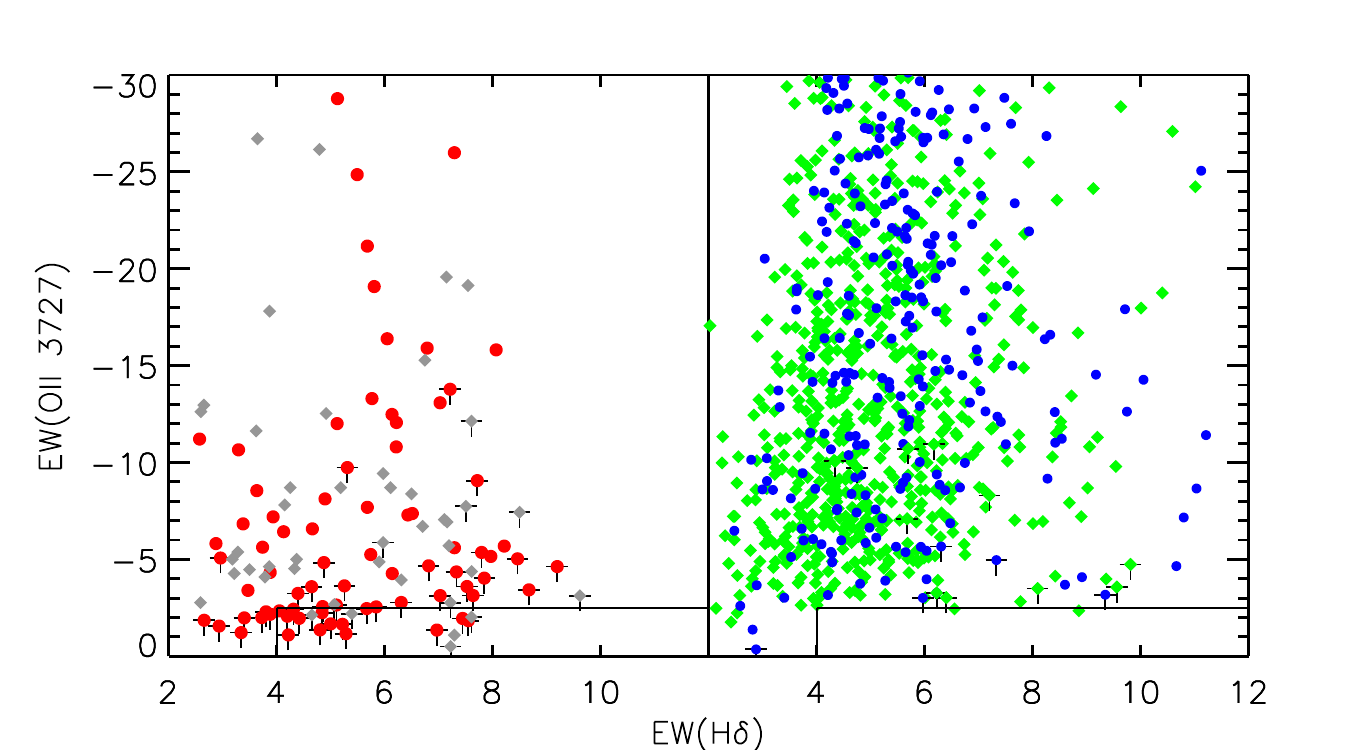}
\caption{H$\delta$ Equivalent width versus [OII] equivalent width for post-starburst selection. The color coding is as same as in above figures. The red and grey dots are galaxies in the red sequence area in the UVJ diagram. The blue and green dots are star forming galaxies.  In this diagram, red galaxies selected in the UVJ diagram have systematically lower [OII] equivalent width than those not in the red sequence area. The solid lines represent EW([OII])$>-$2.5\AA\ and EW(H$\delta$)$>$4\AA, the conventional criteria for post-starburst galaxies. The "$\top$"on the symbol represents the upper limit of EW[OII].}\label{fig:fig8}
\end{figure}

Traditionally Post-starburst galaxies were identified as E+A galaxies or K+A galaxies. The identification process involves either morphology and spectrum or just spectrum alone \citep{2004ApJ...602..190Q,2005MNRAS.359..949B,2005MNRAS.357..937G,2008ApJ...688..945Y,2009MNRAS.398..735Y}. The normal spectroscopic selection for a post-starburst galaxy is to use H$\delta$ and [OII] equivalent width as EW(H$\delta$)$>$4$\sim$6\AA\ and EW([OII] 3727)$>$-2$\sim$-2.5\AA \citep{2005MNRAS.357..937G}. Adopting the criterion, we found that there are 24 objects in our sample with EW(H$\delta$)$>$4\AA\ and EW([OII] 3727)$>$-2.5\AA, shown in Figure~\ref{fig:fig8}. Among these 24 objects, 16 objects are identified in the red sequence in the UVJ diagram in Figure~\ref{fig:fig3} and have log$_{10}$(SFR/SFR$_{\rm MS})<-0.3$ in Figure~\ref{fig:fig6} left panel; 6 are dusty galaxies with MIPS 24$\mu$m detection. On the other hand, there are 53 objects in the red sequence with log$_{10}$(SFR/SFR$_{\rm MS})<-0.3$ and EW(H$\delta$)$>$4\AA. However, most of these galaxies have significant [OII] emission, and some have noisier spectra such that the 3$\sigma$ limit for their [OII] detection is lower than -2.5\AA. \citet{2009MNRAS.398..735Y} argued that [OII] is not a good indicator for star formation. Some elliptical galaxies with weak AGN or LINER in their center can have the [OII] emission line. Additionally, some dusty galaxies with high star formation rates have no [OII] emission due to dust extinction. \citet{2004ApJ...602..190Q} and \citet{2009MNRAS.398..735Y} used spectral decomposition methods by fitting both K- and A-type star spectra to galaxies and determined the A-type star fractions.

Our sample covers a wide redshift range, and the method to identified post-starburst galaxies with H$\delta$ and [OII] EW yields an unfair selection as a function of redshift. Therefore, we propose to use the normalized star formation rate and UVJ colors to weed out post-starburst galaxies. As shown in the right panel of Figure~\ref{fig:fig6}, a galaxy evolving from starburst to a passive galaxy passes through several critical epochs, which could be a criterion. The first point is the peak of EW(H$\delta$) when the evolution track turns around after the starburst era (Figure \ref{fig:fig7}). The turning point for the evolution track indeed changes depending on the star formation history, making it difficult in practice. Another critical epoch could be reaching either log$_{10}$(SFR/SFR$_{\rm MS}$)=0 or log$_{10}$(SFR/SFR$_{\rm MS})$=$-$0.3. Galaxies in this region are no longer on the main sequence, but are transitioning to become passive galaxies, still retaining residual star formation.  The critical line of log$_{10}$(SFR/SFR$_{\rm MS}$)=$-$0.3 is closer to the conventional criterion for post-starburst galaxies. This star formation cut is roughly consistent with the passive galaxies without MIPS 24$\mu$m.

\begin{figure}[ht]
\centering
\includegraphics[width=0.75\linewidth]{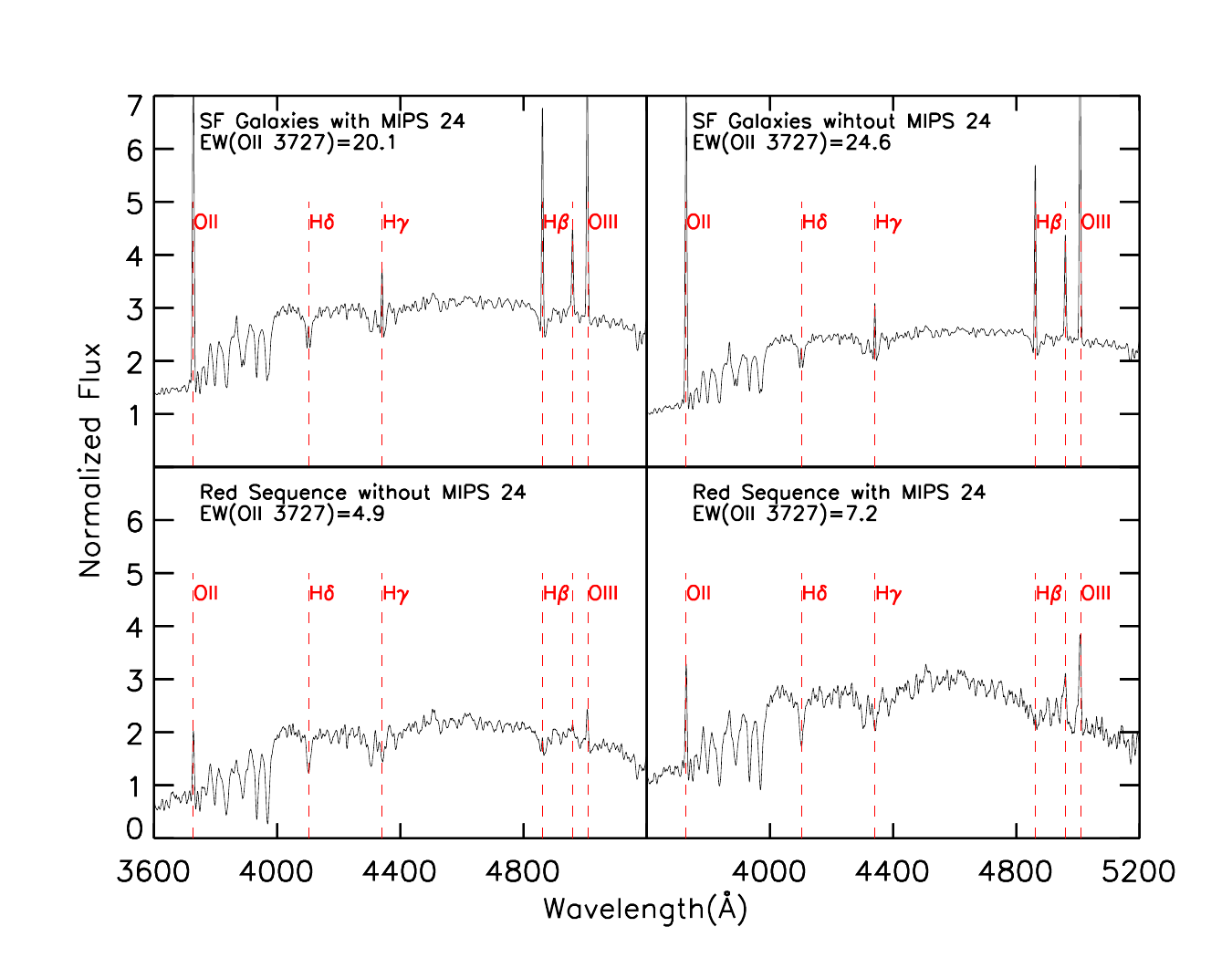}
\caption{Stacked spectrum of galaxies in four populations. Each spectrum had its continuum subtracted before stacking. The continuum-subtracted spectra and their continua were stacked separately, and the resulting stacked spectrum and continuum were then combined. Galaxies in red sequence have significant low [OII] and [OIII] lines lines. Their equivalent width are in Table~2.}
\label{fig:fig9}
\end{figure}

Figure~\ref{fig:fig9} shows stacking spectra with various galaxy populations. The stacking spectrum for galaxies not identified as red sequence in the UVJ diagram, have very strong [OII] and [OIII] emission lines, and H$\beta$, H$\gamma$, and H$\delta$ emission lines superposing in their absorption lines. The passive galaxies without MIPS 24$\mu$m show very weak [OII] and [OIII] emission lines, but no emission line components in H$\beta$, H$\gamma$, and H$\delta$. Passive galaxies with MIPS 24$\mu$m have slightly higher [OII] and [OIII] lines, and we start to detect the H$\beta$ emission line, as shown in Figure~\ref{fig:fig9}.  Therefore, we argue that our selection of EW(H$\delta$) $>$ 4\AA, log$_{10}$(SFR/SFR$_{\rm MS}) <-$0.3, and colors in the red sequence region in the UVJ diagram is very close to the conventional post-starburst selection. In our sample, 53 galaxies satisfy this set of criteria. On the other hand, the conventional spectroscopic selection of post-starburst galaxies (e.g., EW(H$\delta$)$>$4$\sim$6\AA\ and no detectable emission orange of H$\alpha$ and [OII] 3727\AA\ in \citealt{2007MNRAS.381..187G}) may miss substantial amount galaxies with A-type star population and ceasing star formation.

\begin{figure}[ht]
\centering
\includegraphics[width=0.75\linewidth]{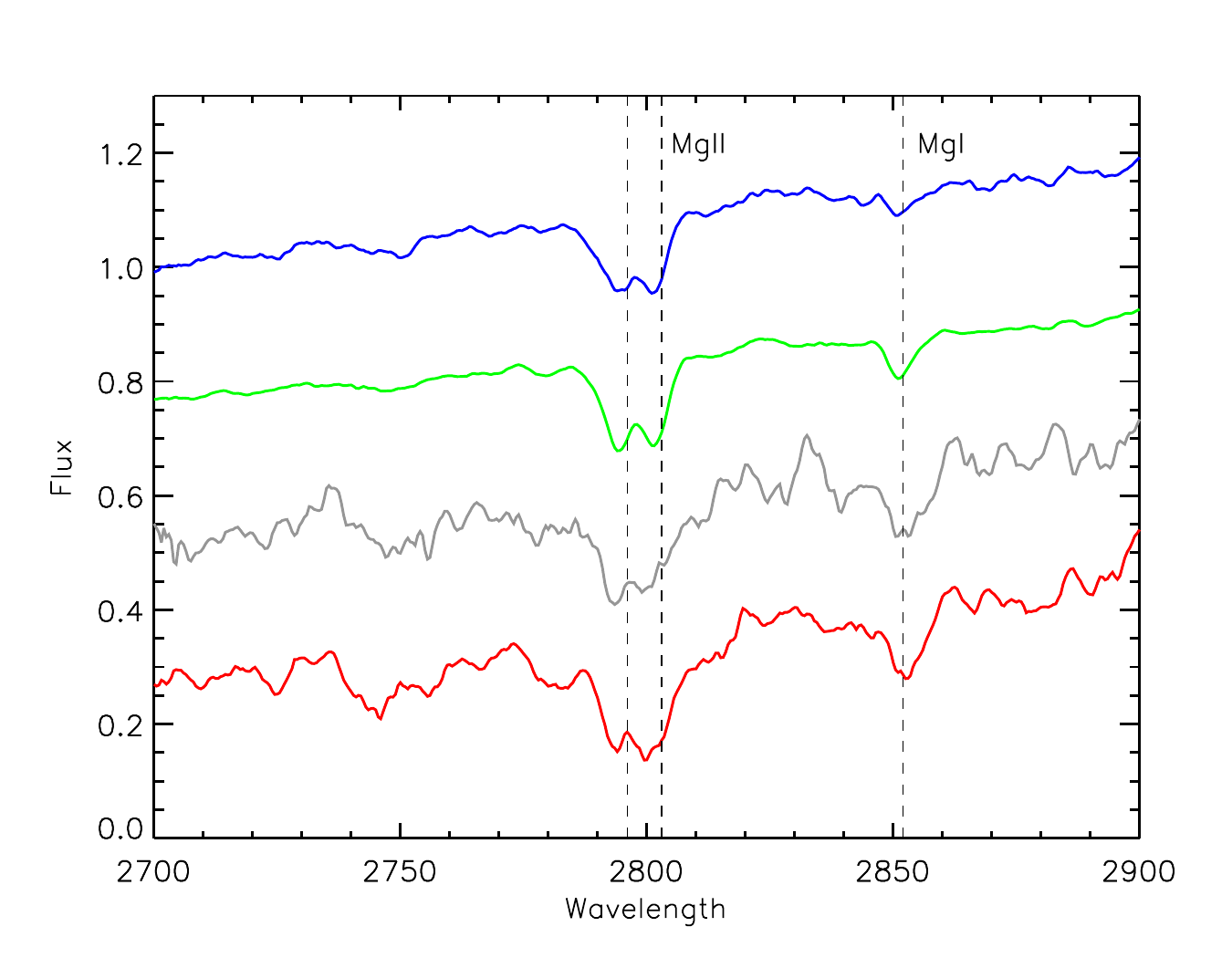}
\caption{Stacked spectrum for galaxies in the 2700$<\lambda<$2900\AA\ range for the MgII lines. Spectra are stacked for galaxies at $z>0.35$, where this wavelength range is available. The galaxies are categorized in the four populations. The color coding is as defined in Figure~\ref{fig:fig3}.}
\label{fig:fig10}
\end{figure}

We also investigated the MgII lines for this sample(Table\ref{tab:2}). \citet{2018A&A...617A..62F} conducted MgII spectroscopy for a sample of star-forming galaxies in the UDF. They found a wide range of MgII profiles, including emission, absorption, and P-Cygni types. However, most of the galaxies in their sample did not show any detection of MgII lines. The MgII emission doublet ratio is consistent with the HII region model, indicating their origin from star-forming regions. We stacked spectra for all galaxies at $z>0.35$ when spectra at the rest-frame $\lambda\sim$ 2800\AA\ are available in their observed optical spectroscopic data. The sample was divided as in Figure~\ref{fig:fig9}. All stacked spectra showed MgI absorption lines at 2852\AA\ and MgII absorption doublet at 2796 and 2803\AA\ (Figure~\ref{fig:fig10}). The passive galaxies without MIPS 24$\mu$m had the largest mean MgII 2800 feature equivalent width of EW(MgII 2800)=7.2\AA. The passive galaxies with MIPS 24$\mu$m had EW(MgII 2800)=6.5\AA. The remaining groups had EW(MgII 2800) around 3\AA. The MgII absorption lines start to appear in B-type stars and reach their maximum in F-G type stars \citep{1990ApJ...364..272F}. The stacking spectra also showed the same trend for the MgI line at 2852\AA, which traces even older stellar populations. Therefore, our selection of passive color in the UVJ and without MIPS 24$\mu$m selects galaxies that already had A-, F-, and possibly G-type stars.

\begin{deluxetable}{lllll}
\linespread{1}
\tablewidth{500pt} 
\tablenum{2}
\tablecaption{Mean line equivalent width for each populations in the Sample. \label{tab:2}}

\startdata 
& & \\
    Population   & MgII(2800)  &MgI(2852) &[OII](3727)&[OIII](5008) \\[6pt]
 \hline   
 red sequence without MIPS 24 detection & 7.2 & 1.3 & -4.9  &-1.6\\ 
 red sequence with MIPS 24 detection & 6.5 & 1.4 & -7.2 & -6.1\\  
 star forming galaxies with MIPS 24 detection & 3.7 & 0.5 & -20.1 & -9.5\\ 
 star forming galaxies without MIPS 24 detection & 2.6 & 0.4 & -24.6 & -13.0\\
\enddata

\end{deluxetable}

\section{AGN Activity in This Sample} \label{sec:AGN}

\begin{figure}[ht]
\centering
\includegraphics[width=0.75\linewidth]{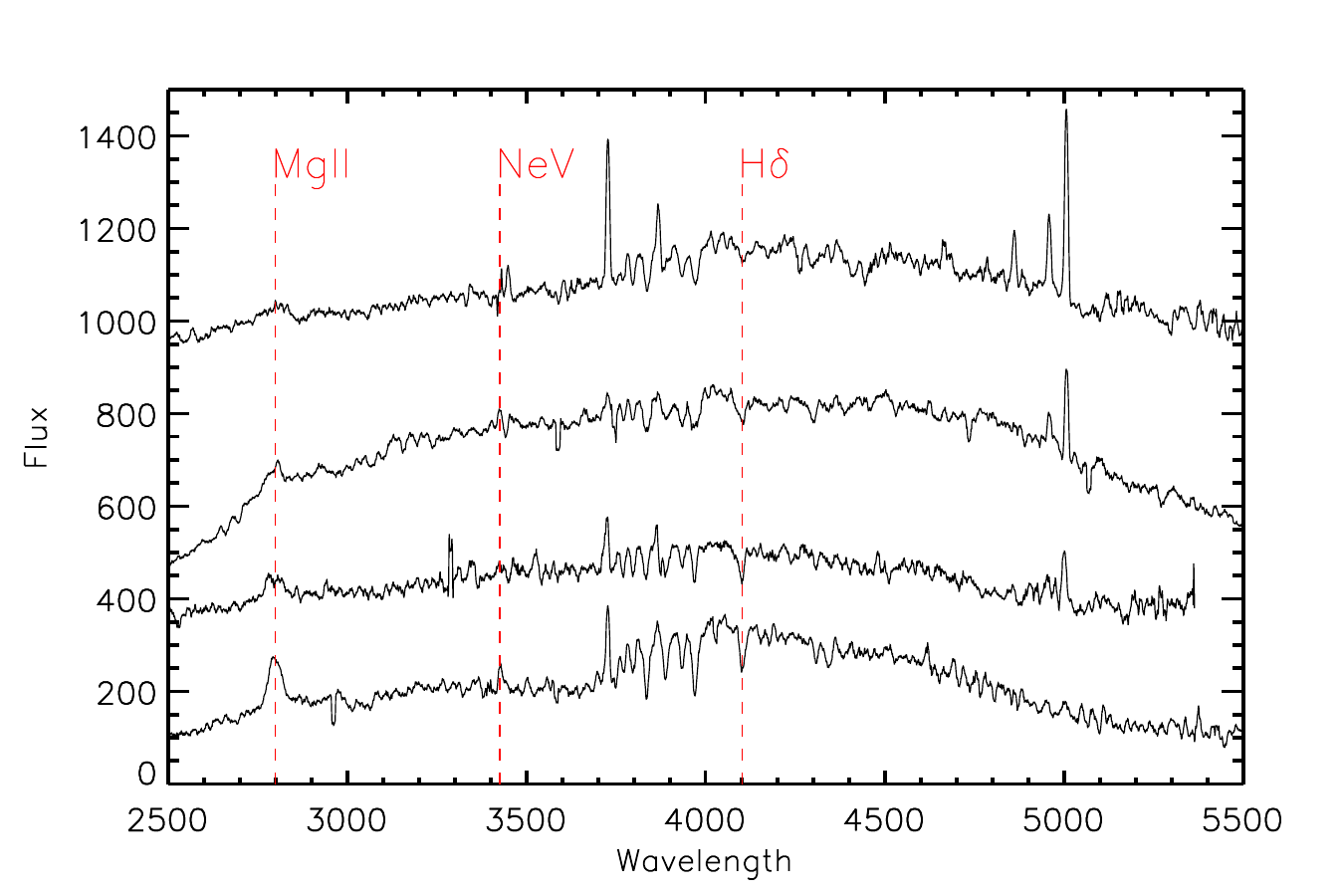}
\caption{Four spectra of galaxies with broad  MgII emission lines. The object with the strongest MgII emission lines does not even have [OIII] line. But three of them also have the [NeV] line.}
\label{fig:fig11}
\end{figure}

AGNs are proposed to play a crucial role in quenching star formation. \citet{2024ApJ...963...99L} used [NeV] emission line at 3426\AA\ to identify AGNs in star forming galaxies. We conducted a similar search for [NeV] emission line and broad MgII emission lines in this sample. We found 61 galaxies with [NeV] emission line at a 3$\sigma$ significance level. Additionally, we searched for broad-line AGNs using MgII and Balmer lines. 
Our selection is biased against Balmer emission lines, only one object in the sample has a broad H$\alpha$ emission line at $z<0.37$. At $z>0.37$, the MgII doublet lines are shifted into optical bands. We found 4 objects with broad MgII lines shown in Figure~\ref{fig:fig11}.

We also identify IR-selected AGNs in the sample using their Mid-IR colors \citet{2007ApJ...660..167D}. Though only one of our IR-color selected AGN has [NeV] line, \citet{2024ApJ...963...99L} used stacking technique to show
 that AGN populations selected with the \citet{2007ApJ...660..167D} criteria exhibit an average [NeV] emission lines at 3426\AA.
Additionally, \citet{2020ApJS..251....2M} conducted a deep wide-field Chandra imaging in the Bo\"{o}tes, revealing X-ray emission with a full band depth of 5$\times$10$^{-16}\, \rm erg \, cm^{-2} \, s^{-1}$. Upon matching our sample to their Chandra catalogue, we found 19 objects with X-ray emission. Their X-ray luminosity falls within the range of $7\times 10 ^{41}<L_{\rm X}< 2\times 10^{43}\, \rm erg \, s^{-1}$, with only two of them having $L_{\rm X}$ slightly smaller than $10^{42}\, \rm erg \, s^{-1}$, which we still consider as AGNs. Three of the X-ray sources have [NeV] emission lines. Two X-ray sources are also identified as IR AGNs with the IR-color selection.

\begin{figure}[ht]
\centering
\includegraphics[width=0.75\linewidth]{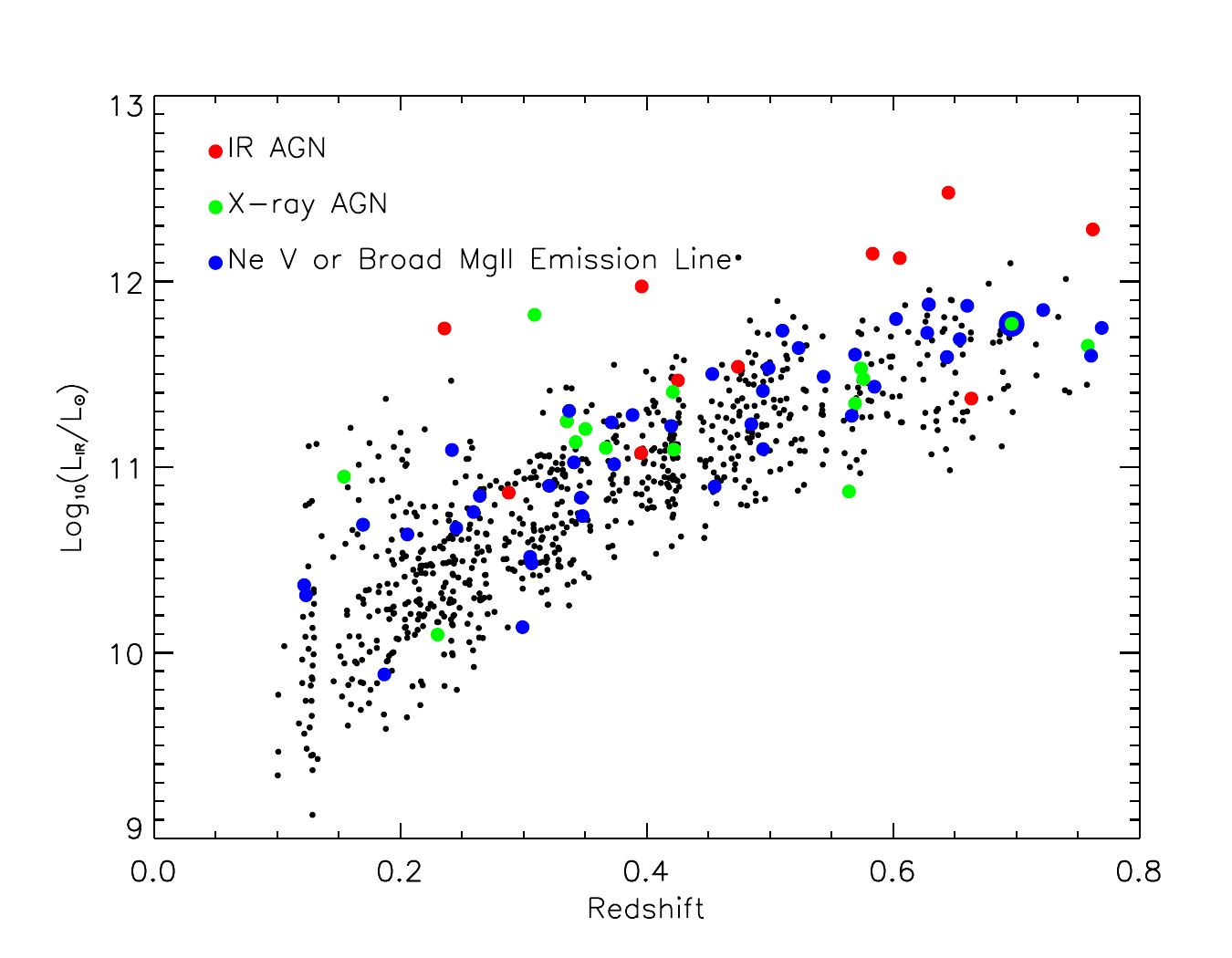}
\caption{
Redshift-Infrared-Luminosity plot illustrating IR, X-ray and broad-line selected AGNs. Infrared Luminosity is defined as the luminosity in the 8$<\lambda<$1000$\mu$m range, derived using Spitzer MIPS 24, 70, Herschel 250, 350, and 500$\mu$m band photometry in the Bo\"{o}tes field \citep{2012MNRAS.424.1614O,2014MNRAS.444.2870W}. A total of 93 IR, X-ray, [NeV], and broad-line selected AGNs are included in the plot.  Five AGNs are ULIRGs, and 49 of them are LIRGs. Most AGNs exhibit higher infrared luminosity in each redshift bin.
}
\label{fig:fig12}
\end{figure}

By adding all X-ray, infrared and spectral selected AGNs together, the total number of AGNs in our sample is 93, constituting approximately 7\% of the sample. Most of these AGNs were found in galaxies with high infrared luminosity, as depicted in Figure~\ref{fig:fig12}. 
In fact, all ULIRGs and the majority of objects with $\log_{10}(L_{\rm IR}/L_{\odot})>$11.5 in this sample are AGNs, indicating the co-existence of the AGN and star formation activity. 
We also examined the UVJ colors for these galaxies with AGNs. About 30\% galaxies in the red sequence area with MIPS 24 micron detection are AGNs, the highest proportion among all populations. 
If AGN-induced quenching occurs, it is likely to happen in this population. 
The remaining AGNs make up 4-8\% of their respective populations in the UVJ color classification summarized in Table 3.

\begin{deluxetable}{lcc}
\tablewidth{500pt} 
\tablenum{3}
\tablecaption{AGNs in the Sample.
}
\label{tab:3}
\startdata 
& & \\
    Population   & Number &  Fraction \\[6pt]
 \hline    
 Red sequence AGNs without MIPS 24 detection & 3 & 4\% \\ 
 Red sequence AGNs with MIPS 24 detection & 13 &  30\%\\  
 AGNs with MIPS 24 detection & 60 & 8\%\\ 
 AGNs without MIPS 24 detection & 17 & 4\%\\
\enddata
\end{deluxetable}

\section{Summary}

In this paper we try to establish an evolution link between star forming and quiescent galaxies by identifying A-type stars signatures in galaxies. 
The appearance of A-type stars, traced by the Balmer absorption lines in a galaxy spectrum, indicates later stage in its star formation history or even suggests that it has already become a post-starburst galaxy.
or even already to become a post-starburst galaxy.
We selected a subset of galaxies from the AGES spectroscopic sample using H$\delta$ absorption line with their EW(H$\delta$)$>$2\AA. There are 1323 galaxies selected in the sample. 

We utilized the UVJ diagram to classify galaxy SED type in the sample. There are 123 passive galaxies identified in the sample, though there are still MIPS24 emission detected in some of these galaxies in the UVJ quiescent region. We further identified only 53 galaxies in the UVJ quiescent region situated below the main sequence area with log$_{10}$(SFR/SFR$_{\rm MS})<-$0.3 and EW(H$\delta$)$>$4\AA. 
The majority of galaxies in the sample are characterized as starburst galaxies located above the main sequence in each redshift bin, and are predominantly LIRGs and ULIRGs. 

Notably, most galaxies in our sample are either in the starburst status above the main sequence or quiescent galaxies below it. A few galaxies in the main sequence are those 
passing through the main sequence zone in their latest evolution from starburst galaxies
to quiescent galaxies. Our spectroscopic study shows that these galaxies passing through the main sequence zone have the maximum fraction of A-type stars. This supports our argument of late stage evolution scenario for this sample.


We also used the spectral features to trace galaxy evolution from starburst to quiescent phase.
The H$\delta$ equivalent width increases when star formation rate for a galaxy decrease from intensive starburst stage to main sequence region where the transition occurred. The further evolution to quiescent stage shows a decreasing of the H$\delta$ equivalent width. The [OII](3727\AA) line show a straight decreasing from starburst to quiescent stage. 
 The passive galaxies selected in our sample display very weak [OII](3727\AA) lines with EW([OII])$\sim$4.9\AA, consistent with the traditional selection criteria for post-starburst galaxies. 
Our further investigation revealed an increasing of the MgII and MgI absorption lines when a galaxy
evolved from starburst to quiescent phase, 

 indicating an increasing fraction of F- and G-type stars in the later stage of starburst and quiescent phases.


In this sample, we also identified only 93 AGNs through infrared color, X-ray detection, and spectral line classification. If the quenching of star formation in these galaxies was driven by AGNs, the small fraction of AGNs suggests a significantly shorter timescale for AGN activities when compared to the lifetime of A-type stars. Majority of AGN host galaxies in this sample are LIRGs and ULIRGs. Thus both AGN activities and intense star formation may occur roughly around the same time. Galaxies in the red sequence area with MIPS 24$\mu$m detection have highest AGN fraction of 30\%,  which is much higher than the average fraction in the whole sample. This indicates an era when AGN quenching occurred. 

\begin{acknowledgments}
The authors would like to appreciate the referee for providing very helpful suggestions. This work is sponsored by the National Key R\&D Program of China for grant No.\ 2022YFA1605300, the National Nature Science Foundation of China (NSFC) grants No. 11933003, 12173045. Additional support came from the Chinese Academy of Sciences (CAS) through a grant to the South America Center for Astronomy (CASSACA) in Santiago, Chile. 

\end{acknowledgments}

%

\vspace{5mm}
\facilities{MMT(Hectospec), CXO, Spitzer, Herschel}






\bibliography{structure}{}
\bibliographystyle{aasjournal}



\end{document}